\documentclass[epsfig,12pt]{article}
\usepackage{epsfig}
\usepackage{graphicx}
\usepackage{hyperref}

\usepackage{array}
\usepackage{amsmath}
\usepackage{amssymb}

\newcommand{\beq}{\begin{equation}}   
\newcommand{\eeq}{\end{equation}}
\newcommand{\beqn}{\begin{eqnarray}}   
\newcommand{\eeqn}{\end{eqnarray}}

\begin{document}
\unitlength = 1mm

\def\de{\partial}
\def\Tr{ \hbox{\rm Tr}}
\def\const{\hbox {\rm const.}}  
\def\o{\over}
\def\im{\hbox{\rm Im}}
\def\re{\hbox{\rm Re}}
\def\bra{\langle}\def\ket{\rangle}
\def\Arg{\hbox {\rm Arg}}
\def\Re{\hbox {\rm Re}}
\def\Im{\hbox {\rm Im}}
\def\diag{\hbox{\rm diag}}


\def\QATOPD#1#2#3#4{{#3 \atopwithdelims#1#2 #4}}
\def\stackunder#1#2{\mathrel{\mathop{#2}\limits_{#1}}}
\def\stackreb#1#2{\mathrel{\mathop{#2}\limits_{#1}}}
\def\Tr{{\rm Tr}}
\def\res{{\rm res}}
\def\Bf#1{\mbox{\boldmath $#1$}}
\def\balpha{{\Bf\alpha}}
\def\bbeta{{\Bf\beta}}
\def\bgamma{{\Bf\gamma}}
\def\bnu{{\Bf\nu}}
\def\bmu{{\Bf\mu}}
\def\bphi{{\Bf\phi}}
\def\bPhi{{\Bf\Phi}}
\def\bomega{{\Bf\omega}}
\def\blambda{{\Bf\lambda}}
\def\brho{{\Bf\rho}}
\def\bsigma{{\bfit\sigma}}
\def\bxi{{\Bf\xi}}
\def\bbeta{{\Bf\eta}}
\def\d{\partial}
\def\der#1#2{\frac{\d{#1}}{\d{#2}}}
\def\Im{{\rm Im}}
\def\Re{{\rm Re}}
\def\rank{{\rm rank}}
\def\diag{{\rm diag}}
\def\2{{1\over 2}}
\def\ntwo{${\mathcal N}=2\;$}
\def\nfour{${\mathcal N}=4\;$}
\def\none{${\mathcal N}=1\;$}
\def\ntwot{${\mathcal N}=(2,2)\;$}
\def\ntwoo{${\mathcal N}=(0,2)\;$}
\def\x{\stackrel{\otimes}{,}}

\newcommand{\cpn}{CP$(N-1)\;$}
\newcommand{\wcpn}{wCP$_{N,\widetilde{N}}(N_f-1)\;$}
\newcommand{\wcpd}{wCP$_{\widetilde{N},N}(N_f-1)\;$}
\newcommand{\vp}{\varphi}
\newcommand{\pt}{\partial}
\newcommand{\tN}{\widetilde{N}}
\newcommand{\ve}{\varepsilon}
\renewcommand{\theequation}{\thesection.\arabic{equation}}

\newcommand{\sun}{SU$(N)\;$}

\setcounter{footnote}0

\vfill

\begin{titlepage}

\begin{flushright}
FTPI-MINN-17/06, UMN-TH-3622/17\\
\end{flushright}

\begin{center}
{  \Large \bf  
 Critical Non-Abelian Vortex in Four
\\[2mm]
 Dimensions  
and Little String Theory
}

\vspace{5mm}

{\large \bf   M.~Shifman$^{\,a}$ and \bf A.~Yung$^{\,\,a,b,c}$}
\end {center}

\begin{center}

$^a${\it  William I. Fine Theoretical Physics Institute,
University of Minnesota,
Minneapolis, MN 55455}\\
$^{b}${\it National Research Center ``Kurchatov Institute'', 
Petersburg Nuclear Physics Institute, Gatchina, St. Petersburg
188300, Russia}\\
$^{c}${\it  St. Petersburg State University,
 Universitetskaya nab., St. Petersburg 199034, Russia}
\end{center}

\vspace{1cm}

\begin{center}
{\large\bf Abstract}
\end{center}

As was shown recently,  non-Abelian vortex strings supported in four-dimensional 
\ntwo supersymmetric QCD with the U(2) gauge group
and $N_f=4$  quark multiplets (flavors) become critical superstrings. In addition to the
translational moduli non-Abelian strings under consideration carry six orientational and size moduli. 
Together they form a ten-dimensional target space required for a superstring to 
be critical.  The  target space of the string sigma model is a product of the flat four-dimensional
space and a Calabi-Yau non-compact threefold, namely, the conifold.
We study closed string states which emerge in four dimensions (4D) and identify them with
 hadrons of four-dimensional  \ntwo QCD. One massless  state was found previously: it  emerges as
a massless hypermultiplet associated with the deformation of the complex structure of the conifold. 
In this paper we find a number of massive states. To this end we exploit the approach used in LST, ``Little
String Theory,'' namely equivalence between the 
critical string  on the conifold and non-critical $c=1$ string with the Liouville field and a compact scalar at 
the self-dual radius. The states we find carry ``baryonic'' charge (its definition differs from standard). We
 interpret them as ``monopole necklaces'' formed (at strong coupling) by the closed string with 
confined monopoles attached.

\vspace{2cm}

\end{titlepage}

\newpage


\newpage

\section {Introduction }
\label{intro}
\setcounter{equation}{0}

It was recently shown \cite{SYcstring} that the non-Abelian solitonic vortex string in a 
certain 4D Yang-Mills theory becomes critical  at strong coupling.
This particular 4D theory in which the non-Abelian vortex is critical is \ntwo supersymmetric QCD with
the U(2) gauge group, four quark flavors and the
Fayet-Iliopoulos (FI) \cite{FI} parameter $\xi$.  

Non-Abelian vortices were first discovered in 
\ntwo supersymmetric QCD with the gauge group U$(N)$ and $N_f \ge N$ flavors of quark hypermultiplets
\cite{HT1,ABEKY,SYmon,HT2}.
The non-Abelian vortex string is 1/2
BPS saturated and, therefore,  has \ntwot supersymmetry on its world sheet.
In addition to four translational moduli characteristic of the  Abrikosov-Nielsen-Olesen (ANO) strings 
\cite{ANO}, the non-Abelian string carries orientational  moduli, as well as the size moduli if $N_f>N$
\cite{HT1,ABEKY,SYmon,HT2} (see \cite{Trev,Jrev,SYrev,Trev2} for reviews). Their dynamics
are described by effective two-dimensional sigma model on the string world sheet with 
the target space 
\beq
\mathcal{O}(-1)^{\oplus(N_f-N)}_{\mathbb{CP}^1}\,,
\label{12}
\eeq
to which we will refer to as to the weighted CP  model (WCP$(N,N_f-N)$). 
 For $N_f=2N$
the model becomes conformal. Moreover for $N=2$ the 
dimension of the orientational/size moduli space is six and they can be combined with 
four translational moduli to form a ten-dimensional space required for 
superstring to become critical.\footnote{This set up corresponds to the Virasoro central charge
in the WCP$(N,N_f-N)$ sector
$c_{\rm WCP}=9$.} In this case the target space of the world sheet 2D theory on 
the non-Abelian vortex string is
 $\mathbb{R}^4\times Y_6$, where $Y_6$ is a non-compact six dimensional Calabi-Yau manifold, the 
resolved conifold \cite{Candel,NVafa}.

The main obstacle in describing the solitonic vortex string as a critical string is that
the solitonic strings are typically thick. Their transverse size is given  by $1/m$, where $m$ is 
a typical mass scale 
of the four-dimensional fields forming the string. This leads to the presence of a series of higher derivative
corrections to the low-energy sigma model action. The higher derivative corrections run in powers of 
$\partial/m$. They make the string world sheet ``crumpled'' \cite{Polyak86} and the string does not
produce linear Regge trajectories at small spins \cite{SYcstring}.

The higher derivative corrections on the non-critical string world sheet are  needed
to be properly accounted for in order
 to improve the ultra-violet (UV)
behavior of the string theory \cite{PolchStrom}. Without them the low-energy world sheet sigma model
does not lead to UV complete string theory. In particular, this means that, say, the ANO string
in four dimensions \cite{ANO} never becomes thin.

On the other hand, the non-Abelian vortex string on the conifold is critical and has 
a perfectly good UV
behavior. This opens the possibility that it can become thin in a certain regime. This cannot happen in weakly
coupled bulk
 theory because at weak coupling $m\sim g\sqrt{T}$ and is always small in the units of $\sqrt{T}$. 
 Here  $g$ is the gauge coupling constant 
of the four-dimensional ${\cal N}=2$  QCD and $T$ is  the string tension.

A conjecture was put forward in  \cite{SYcstring}  that  at strong coupling 
in the vicinity of a critical value of $g_c^2\sim 1$ the non-Abelian string on the conifold becomes thin,
and higher-derivative corrections in the action can be ignored. It is expected that
the thin string produces linear Regge trajectories for all spins.
 The above  conjecture implies\,\footnote{At $N_f=2N$ the beta function
of the 4D \ntwo QCD is zero, so the gauge coupling $g^2$ does not run. Note, however, that conformal
invariance in the 4D theory is broken by the FI parameter $\xi$ which does not run either.} that $m(g^2) \to \infty$ at  $ g^2\to g_c^2$. Moreover, it was argued in 
\cite{KSYcstring,KSYconifold} that it is natural to expect that the critical point $g_c$ where the vortex string
becomes thin is the self-dual point $g_c^2=4\pi$, see \cite{ArgPlessShapiro,APS}.

A version of the string-gauge duality
for 4D QCD  was proposed \cite{SYcstring}: at weak coupling this 
theory is in the Higgs phase and can be 
described in terms of (s)quarks and Higgsed gauge bosons, while at strong coupling hadrons of this theory 
can be understood as string states formed by the non-Abelian vortex string.
This hypothesis was further explored by studying string theory for the critical 
non-Abelian vortex in \cite{KSYcstring,KSYconifold}. 

The vortices in the U$(N)$ theories under consideration
are topologically stable and cannot be broken. Therefore
the finite length strings are closed. Thus, we focus on the 
closed strings. The goal is to  identify closed string states 
with  hadrons of the 4D \ntwo QCD. 

The first
step of this program, namely,  identifying massless
string states was carried out in \cite{KSYcstring,KSYconifold} using supergravity formalism.
In particular,  a single matter hypermultiplet associated with the deformation 
of the complex structure of the conifold was found as the only 4D {\em massless} mode of the string. 
Other states arising from the massless ten-dimensional graviton are not dynamical
in four dimensions. In particular, the 4D graviton and  unwanted vector multiplet associated with
deformations of the K\"ahler form of the conifold are absent.
This is due to non-compactness of the  Calabi-Yau manifold we deal with and 
non-normalizability of the corresponding  modes over six-dimensional space $Y_6$.

Moreover, it was also discussed \cite{KSYcstring,KSYconifold} how the states seen in 4D \ntwo  QCD at weak coupling are related to
what we obtain from the string theory at strong coupling. In particular,  the
 hypermultiplet associated with the deformation of the complex structure  of the conifold was interpreted 
as a monopole-monopole baryon \cite{KSYcstring,KSYconifold}.

In this paper we make the next step and find a number of massive states of the closed non-Abelian vortex 
string which we interpret as hadrons of 4D \ntwo QCD.  However, to this end we cannot use 
our formulation of the critical string theory on the conifold. The point is that
the coupling constant $1/\beta$ of the world sheet WCP(2,2)  is not small. Moreover $\beta$ tends to zero
once the 4D coupling $g^2$ approaches the self-dual value we are interested in. At $\beta\to 0$ the resolved
conifold develops a conical singularity. The supergravity approximation does not work 
for massive states.\footnote{This is in contradistinction to the massless states. 
For the latter, we can perform
computations at large $\beta$ where the supergravity approximation is valid and 
then extrapolate to strong coupling.
In the sigma-model language this procedure corresponds to chiral primary operators. They are protected by
\ntwot world-sheet supersymmetry and their masses are not lifted by quantum corrections.}

To analyze the massive states we apply a different approach
which was used for Little String Theories (LST), see  \cite{Kutasov} for a review.
Namely, we use  the equivalence between the 
critical string  on the conifold and non-critical $c=1$ string which contains the Liouville 
field and a compact scalar at 
the self-dual radius \cite{GivKut,GVafa}. The latter theory (in the WZNW formulation) can be
analyzed by virtue of algebraic methods.  
The spectrum  can be computed exactly \cite{DixonPeskinLy,Petrop,Hwang,MukVafa,EGPerry}.

The states which we find carry a ``baryonic'' charge -- we
 interpret them  as ``monopole necklaces'' formed (at strong coupling) by the closed string with 
confined monopoles attached.

It is worth mentioning  that the solitonic vortex describes only non-pertur\-bative states. 
Perturbative states, in particular massless states associated with the Higgs 
branch in the original 4D Yang-Mills theory are present for all values of the gauge coupling and 
are not captured by the vortex string dynamics. 

The paper is organized as follows. In Sec. \ref{worldsheet} we review the world-sheet sigma model
emerging on the critical non-Abelian vortex string. In Sec. \ref{conifold} we briefly review conifold geometry
and the massless state associated with deformations of the conifold complex structure. In Sec. \ref{c=1}
we describe the equivalent formulation in terms of the non-critical $c=1$ string, and in Sec. \ref{spectrum}
we calculate its spectrum. Section \ref{necklace} presents an interpretation of  states we found in 
terms of the baryonic ``monopole necklaces.'' We summarize our conclusions in Sec. \ref{conclusions}.

\section {Non-Abelian vortex string }
\label{worldsheet}
\setcounter{equation}{0} 

\subsection{Four-dimensional \ntwo QCD}

As was already mentioned non-Abelian vortex-strings were first found in 4D
\ntwo supersymmetric QCD with the gauge group U$(N)$ and $N_f \ge N$ flavors of the quark hypermultiplets
supplemented by the FI $D$ term $\xi$
\cite{HT1,ABEKY,SYmon,HT2}, see for example \cite{SYrev} for a detailed review of this theory.
Here we just mention that at weak coupling $g^2\ll 1$ this theory is in the Higgs phase where scalar
components of quark multiplets (squarks) develop vacuum expectation values (VEVs). These VEVs breaks 
the U$(N)$ gauge group
Higgsing {\em all} gauge bosons, while the global flavor SU$(N_f)$ is broken down to the so called color-flavor
locked group. The resulting global symmetry is
\beq
 {\rm SU}(N)_{C+F}\times {\rm SU}(N_f-N)\times {\rm U}(1)_B,
\label{c+f}
\eeq
see \cite{SYrev} for more details. The unbroken global U(1)$_B$ factor above is identified with a baryonic symmetry. Note that 
what is usually identified as the baryonic U(1) charge is a part of  our 4D theory  gauge group.
 ``Our" U(1)$_B$
is  an unbroken by squark VEVs combination of two U(1) symmetries:  the first is a subgroup of the flavor 
SU$(N_f)$ and the second is the global U(1) subgroup of U$(N)$ gauge symmetry.

The 4D theory has a Higgs branch formed by massless quarks which are in  the bifundamental representation
of the global group \eqref{c+f} and carry baryonic charge, see \cite{KSYconifold} for more details.
In the case $N=2$, $N_f=2N=4$ we will deal with here the dimension of this branch is 
\beq
{\rm dim}\,{\cal H}= 4N (N_f-N)=16.
\label{dimH}
\eeq
The above Higgs branch is non-compact and is hyper-K\"ahlerian \cite{SW2,APS}, therefore 
its metric cannot be 
modified by quantum corrections \cite{APS}. In particular, once the Higgs branch is present at weak coupling
we can continue it all the way into strong coupling. 

\subsection{World-sheet sigma model}

The presence of color-flavor locked group SU$(N)_{C+F}$ is the reason for the formation of the
non-Abelian vortex strings \cite{HT1,ABEKY,SYmon,HT2}.
The most important feature of these vortices is the presence of the so-called orientational  zero modes.
In the \ntwo 4D theory these strings are 1/2 BPS-saturated; hence,  their
tension  is determined  exactly by the FI parameter,
\beq
T=2\pi \xi\,.
\label{ten}
\eeq

Let us briefly review the model emerging on the world sheet
of the non-Abelian critical string \cite{SYcstring,KSYcstring,KSYconifold}.

The translational moduli fields (they decouple from all other moduli)
 in the Polyakov formulation \cite{P81} are given by the action
\beq
S_{\rm 0} = \frac{T}{2}\,\int d^2 \sigma \sqrt{h}\, 
h^{\alpha\beta}\pt_{\alpha}x^{\mu}\,\pt_{\beta}x_{\mu}
+\mbox{fermions}\,,
\label{s0}
\eeq
where $\sigma^{\alpha}$ ($\alpha=1,2$) are the world-sheet coordinates, $x^{\mu}$ ($\mu=1,...,4$) 
describing the $\mathbb{R}^4$ part  of the string
target space and $h={\rm det}\,(h_{\alpha\beta})$, where $h_{\alpha\beta}$ is the world-sheet metric 
which is understood as an
independent variable.

If $N_f=N$  the dynamics of the orientational zero modes of the non-Abelian vortex, which become 
orientational moduli fields 
 on the world sheet, is described by two-dimensional
\ntwot supersymmetric ${\rm CP}(N-1)$ model.

If one adds extra quark flavors, non-Abelian vortices become semilocal.
They acquire size moduli \cite{AchVas}.  
In particular, for the non-Abelian semilocal vortex at hand,  in 
addition to  the orientational zero modes  $n^P$ ($P=1,2$), there are  the so-called size moduli   
$\rho^K$ ($K=1,2$) \cite{AchVas,HT1,HT2,SYsem,Jsem,SVY}.  

The  gauged formulation of the effective world sheet theory for the orientational and size moduli is as follows \cite{W93}. One introduces
 the U$(1)$ charges $\pm 1$, namely $+1$ for $n$'s and $-1$ for $\rho$'s, 
\beqn
S_{\rm 1} &=& \int d^2 \sigma \sqrt{h} \left\{ h^{\alpha\beta}\left(
 \tilde{\nabla}_{\alpha}\bar{n}_P\,\nabla_{\beta} \,n^{P} 
 +\nabla_{\alpha}\bar{\rho}_K\,\tilde{\nabla}_{\beta} \,\rho^K\right)
 \right.
\nonumber\\[3mm]
&+&\left.
 \frac{e^2}{2} \left(|n^{P}|^2-|\rho^K|^2 -\beta\right)^2
\right\}+\mbox{fermions}\,,
\label{wcp}
\eeqn
where 
\beq 
\nabla_{\alpha}=\pt_{\alpha}-iA_{\alpha}\,, \qquad \tilde{\nabla}_{\alpha}=\pt_{\alpha}+iA_{\alpha}
\eeq
 and $A_\alpha$ is an auxiliary gauge field.
 The limit $e^2\to\infty$ is implied. Equation (\ref{wcp}) represents the   ${\rm WCP}(2,2) $ model.\footnote{Both the orientational and the size moduli
have logarithmically divergent norms, see e.g.  \cite{SYsem}. After an appropriate infrared 
regularization, logarithmically divergent norms  can be absorbed into the definition of 
relevant two-dimensional fields  \cite{SYsem}.
In fact, the world-sheet theory on the semilocal non-Abelian string is 
not exactly the ${\rm WCP}(N,\tN)$ model \cite{SVY}, there are minor differences. The actual theory is called the $zn$ model. Nevertheless it has the same infrared physics as the model (\ref{wcp}) \cite{KSVY}.} 

The total number of real bosonic degrees of freedom in (\ref{wcp}) is six, 
where we take into account the $D$-term constraint
and the fact that one U(1) phase can be gauged away. As was already mentioned, 
these six internal degrees of freedom 
are combined with the four translational moduli from (\ref{s0}) to form a ten-dimensional space needed 
for superstring to be critical.

In the semiclassical approximation the coupling constant $\beta$ in (\ref{wcp}) is related to 
the 4D SU(2) gauge coupling 
$g^2$ via \cite{SYrev}
\beq
\beta\approx \frac{4\pi}{g^2}\,.
\label{betag}
\eeq
Note that the first (and the only) coefficient is the same for the 4D QCD and the
world-sheet beta 
functions. Both vanish at $N_f=2N$. This ensures that our world-sheet theory is conformal.

The total bosonic world-sheet action is
\beq
S=S_0+S_1\,.
\label{stringaction}
\eeq
Since non-Abelian  vortex string  is 1/2 BPS it preserves ${\mathcal N} =(2,2)$  in the world sheet
sigma model which 
is necessary to have \ntwo space-time supersymmetry \cite{Gepner,BDFM}. Moreover, in \cite{KSYconifold}
it is shown that the string theory of the non-Abelian critical vortex is type IIA.

The global symmetry of the world-sheet sigma model (\ref{wcp}) 
is
\beq
 {\rm SU}(2)\times {\rm SU}(2)\times {\rm U}(1)\,,
\label{globgroup}
\eeq
i.e. exactly the same as the unbroken global group in the 4D theory (\ref{c+f}) at  $N=2$ and $N_f=4$. 
The fields $n$ and $\rho$ 
transform in the following representations:
\beq
n:\quad (\textbf{2},\,0,\, 0), \qquad \rho:\quad (0,\,\textbf{2},\, 1)\,.
\label{repsnrho}
\eeq

\subsection{Thin string regime}

As well known \cite{ArgPlessShapiro,APS}  the 4D Yang-Mills theory at hand possesses a strong-weak coupling duality, namely,
\beq
\tau\to \tau_{D} = -\frac{1}{\tau}\,,\qquad \tau = i\frac{4\pi}{g^2} +\frac{\theta_{4D}}{2\pi}\,,
\label{bulkduality}
\eeq 
 where $\theta_{4D}$ is the four-dimensional $\theta$ angle.

The 2D coupling constant $\beta$ can be naturally complexified too if we include the $\theta$ term in the 
action of the  model \eqref{wcp}, 
$$\beta \to \beta + i\,\frac{\theta_{2D}}{2\pi}\,.$$ 
The exact relation between 4D and 2D couplings is as follows:
\beq
\exp{(-2\pi\beta)} = - h(\tau)[h(\tau) +2],
\label{taubeta}
\eeq
where the function $h(\tau)$ is a special modular function of $\tau$ defined in terms of the
$\theta$-functions, $$h(\tau)=\theta_1^4/(\theta_2^4-\theta_1^4)\,.$$
 This function enters the Seiberg-Witten curve for our 4D theory
 \cite{ArgPlessShapiro,APS}. The equation (\ref{taubeta}) generalizes the quasiclassical relation
\eqref{betag}.
Derivation of the relation (\ref{taubeta}) will be presented elsewhere \cite{Komarg,tobepub}.

Note, that the 4D self-dual point $g^2=4\pi$ is mapped onto the 2D self-dual point $\beta=0$.

According to the hypothesis formulated in  \cite{SYcstring}, our critical non-Abelian string becomes thin
in the strong coupling limit in the self-dual point $\tau_c =i$ or $g^2_c =4\pi$ .  This gives
\beq
m^2  \to \xi\times 
\left\{
\begin{array}{ccc}
g^2, & g^2\ll 1& \\[1mm]
\infty, & g^2\to 4\pi& \\[1mm]
16\pi^2/g^2,& g^2\gg 1 & \\
\end{array},
\right.
\label{msing}
\eeq
where the dependence of $m$  at small and large $g^2$ follows from the
quasiclassical analysis \cite{SYrev} and
duality (\ref{bulkduality}),  respectively.

Thus we expect that the  singularity of mass $m$ lies at $\beta=0$. This  is the point where 
the non-Abelian string 
becomes infinitely thin,  higher derivative terms can be neglected and the theory of the non-Abelian 
string reduces to (\ref{stringaction}). The point $\beta=0$ is a natural choice because at this point
 we have a regime change in the 2D sigma model  {\em per se}. 
This is the point where the resolved conifold defined by the $D$ term in
(\ref{wcp}) develops a conical singularity \cite{NVafa}.

\section {Massless 4D baryon as deformation of the conifold complex structure}
\label{conifold}
\setcounter{equation}{0} 

In this section we briefly review the only 4D massless state associated 
with the deformation of the conifold complex structure. It was found in \cite{KSYconifold}.
 As was already mentioned, all other modes arising from massless 10D
graviton have non-normalizable wave functions over the conifold. In particular, 4D graviton is
absent \cite{KSYconifold}. This result matches our expectations since we started with
\ntwo QCD in the flat four-dimensional space without gravity.

The target space of the sigma model (\ref{wcp}) is defined by the $D$-term condition
\beq
|n^P|^2-|\rho^K|^2 = \beta\,.
\label{Fterm}
\eeq 
The U(1) phase is assumed to be gauged away. 
We can construct the U(1) gauge-invariant ``mesonic'' variables
\beq
w^{PK}= n^P \rho^K.
\label{w}
\eeq
These variables are subject to the constraint
${\rm det}\, w^{PK} =0$, or
\beq
\sum_{\alpha =1}^{4} w_{\alpha}^2 =0,
\label{coni}
\eeq
where $$w^{PK}=\sigma_{\alpha}^{PK}w_{\alpha}\,,$$ and the $\sigma$ matrices above
are  $(1,-i\tau^a)$, $a=1,2,3$.
Equation (\ref{coni}) defines the conifold $Y_6$.  
It has the K\"ahler Ricci-flat metric and represents a non-compact
 Calabi-Yau manifold \cite{Candel,W93,NVafa}. It is a cone which can be parametrized 
by the non-compact radial coordinate 
\beq
\widetilde{r}^{\, 2} =\sum_{\alpha =1}^{4} |w_{\alpha}|^2\,
\label{tilder}
\eeq
and five angles, see \cite{Candel}. Its section at fixed $\widetilde{r}$ is $S_2\times S_3$.

At $\beta =0$ the conifold develops a conical singularity, so both $S_2$ and $S_3$  
can shrink to zero.
The conifold singularity can be smoothed out
in two distinct ways: by deforming the K\"ahler form or by  deforming the 
complex structure. The first option is called the resolved conifold and amounts to introducing 
a non-zero $\beta$ in (\ref{Fterm}). This resolution preserves 
the K\"ahler structure and Ricci-flatness of the metric. 
If we put $\rho^K=0$ in (\ref{wcp}) we get the $CP(1)$ model with the $S_2$ target space
(with the radius $\sqrt{\beta}$).  
The resolved conifold has no normalizable zero modes. 
In particular, 
the modulus $\beta$  which becomes a scalar field in four dimensions
 has non-normalizable wave function over the 
$Y_6$ \cite{KSYconifold}.  

As  explained in \cite{GukVafaWitt,KSYconifold}, non-normalizable 4D modes can be 
interpreted as (frozen) 
coupling constants in the 4D  theory. 
The $\beta$ field is the most straightforward example of this, since the 2D coupling $\beta$ is
 related to the 4D coupling, see Eq. (\ref{taubeta}).

If $\beta=0$ another option exists, namely a deformation 
of the complex structure \cite{NVafa}. 
It   preserves the
K\"ahler  structure and Ricci-flatness  of the conifold and is 
usually referred to as the {\em deformed conifold}. 
It  is defined by deformation of Eq.~(\ref{coni}), namely,   
\beq
\sum_{\alpha =1}^{4} w_{\alpha}^2 = b\,,
\label{deformedconi}
\eeq
where $b$ is a complex number.
Now  the $S_3$ can not shrink to zero, its minimal size is 
determined by
$b$. 

The modulus $b$ becomes a 4D complex scalar field. The  effective action for  this field was calculated in \cite{KSYconifold}
using the explicit metric on the deformed conifold  \cite{Candel,Ohta,KlebStrass},
\beq
S(b) = T\int d^4x |\pt_{\mu} b|^2 \,
\log{\frac{T^2 L^4}{|b|}}\,,
\label{Sb}
\eeq
where $L$ is the  size of $\mathbb{R}^4$ introduced as an infrared regularization of 
logarithmically divergent $b$ field 
norm.\footnote{The infrared regularization
on the conifold $\widetilde{r}_{\rm max}$ translates into the size $L$ of the 4D space 
 because variables  $\rho$ in \eqref{tilder} have an interpretation of the vortex string sizes,
$\widetilde{r}_{\rm max}\sim TL^2$ .}

We see that the norm of
the $b$ modulus turns out to be  logarithmically divergent in the infrared.
The modes with the logarithmically divergent norm are at the borderline between normalizable 
and non-normalizable modes. Usually
such states are considered as ``localized'' on the string. We follow this rule.  We can
 relate this logarithmic behavior to the marginal stability of the $b$ state, see \cite{KSYconifold}.
This scalar mode is localized on the string in the same sense as the orientational 
and size zero modes are localized on the vortex-string solution.
   
 The field $b$  being massless can develop a VEV. Thus, 
we have a new Higgs branch in 4D \ntwo QCD which is developed only for the self-dual value of 
coupling constant $g^2=4\pi$. 

The logarithmic metric in (\ref{Sb}) in principle can receive both perturbative and 
non-perturbative quantum corrections in the sigma model coupling $1/\beta$. However, for 
\ntwo  theory the non-renormalization
theorem of \cite{APS} forbids the  dependence of the Higgs branch metric  on the 4D coupling 
constant $g^2$.
Since the 2D coupling $\beta$ is related to $g^2$ we expect that the logarithmic metric in (\ref{Sb})
will stay intact. We confirm this expectation in the next section.

 In \cite{KSYconifold} the massless state $b$ was interpreted as a baryon of 4D \ntwo QCD.
Let us explain this.
 From Eq.~(\ref{deformedconi}) we see that the complex 
parameter $b$ (which is promoted to a 4D scalar field) is singlet with respect to both SU(2) factors in
 (\ref{globgroup}), i.e. 
the global world-sheet group.\footnote{Which is isomorphic to the 4D
global group \eqref{c+f} at $N=2$, $N_f=4$.} What about its baryonic charge? 

Since
\beq
w_{\alpha}= \frac12\, {\rm Tr}\left[(\bar{\sigma}_{\alpha})_{ KP}\,n^P\rho^K\right]
\label{eq:kinkbaryon}
\eeq
we see that the $b$ state transforms as 
\beq
(1,\,1,\,2),
\label{brep}
\eeq
where we used  (\ref{repsnrho}) and (\ref{deformedconi}). In particular it has the baryon charge $Q_B(b)=2$.

To conclude this section let us note that in type IIA superstring the complex scalar 
associated with deformations of the complex structure of the Calabi-Yau
space enters as a 4D \ntwo hypermultiplet. Other components of this hypermultiplet can be restored 
by \ntwo supersymmetry. In particular, 4D \ntwo hypermultiplet should contain another complex scalar $\tilde{b}$
with baryon charge  $Q_B(\tilde{b})=-2$. In the stringy description this scalar comes from ten-dimensional
three-form, see \cite{Louis} for a review.

\section {Non-critical \boldmath{$c=1$} string }
\label{c=1}
\setcounter{equation}{0} 

As was explained in Sec. \ref{intro} the critical string theory on the conifold is hard
to use for calculating the spectrum of massive string modes because the supergravity approximation
does not work. In this paper we take a different route and use the equivalent formulation of our
 theory as a non-critical   $c=1$ string theory with the Liouville field and a compact scalar at 
the self-dual radius \cite{GivKut,GVafa}.

Non-critical $c=1$ string theory is formulated on the target space
\beq
\mathbb{R}^4\times \mathbb{R}_{\phi}\times S^1,
\label{target}
\eeq
where $\mathbb{R}_{\phi}$ is a real line associated with the Liouville field $\phi$ and the theory  
has a linear in $\phi$ dilaton, such that string coupling is given by
\beq
g_s =e^{-\frac{Q}{2}\phi}\, .
\label{strcoupling}
\eeq

Generically the above  equivalence is formulated between the critical string on non-compact Calabi-Yau spaces with 
isolated singularity on the one hand, and non-critical $c=1$ string with the additional Ginzburg-Landau
\ntwo superconformal system \cite{GivKut} on the other hand. In  the conifold case  this extra Ginzburg-Landau factor in \eqref{target} is absent \cite{GivKutP}.

 In \cite{ABKS,GivKut,GivKutP} it was argued that non-critical string theories with the string
coupling exponentially falling off at $\phi\to\infty$ are holographic. The string coupling
goes to zero in the bulk of the  space-time  and non-trivial dynamics (LST)\,\footnote{The main 
example of this behavior is non-gravitational LST
in the flat six-dimensional space formed by the world volume of parallel NS5 branes.}  is 
localized on the ``boundary.'' 
In our case the ``boundary'' is the four-dimensional space in which \ntwo QCD is defined.

The holography for our non-Abelian vortex string theory is most welcome and expected. We start with 
\ntwo QCD in 4D space and study solitonic vortex string. In our approach 10D space formed by 4D ``real''
space and six internal moduli of the string is an artificial construction needed to formulate the string
theory of a special non-Abelian vortex. Clearly we expect that all non-trivial ``real" physics should be localized exclusively on the 
4D ``boundary.'' In other words,  we expect that LST in our case is nothing other than 4D
\ntwo supersymmetric QCD at the self-dual value of the gauge coupling $g^2=4\pi$ (in the hadronic description).

The linear dilaton in \eqref{strcoupling} implies that the bosonic stress tensor of $c=1$ matter coupled to 
2D gravity is given by
\beq
T_{--}= -\frac12\,\left[(\pt_z \phi)^2 + Q\, \pt_z^2 \phi + (\pt_z Y)^2\right]. 
\label{T--}
\eeq
The  compact scalar $Y$  represents $c=1$ matter and satisfies the following condition:
\beq
Y \sim Y+2\pi Q\, .
\eeq
Here we normalize the scalar fields in such a way that their propagators are
\beq
\langle \phi(z),\phi(0)\rangle = -\log{z\bar{z}}, \qquad \langle Y(z),Y(0)\rangle = -\log{z\bar{z}}\,.
\label{propagators}
\eeq
The central charge of the supersymmetrized $c=1$ theory above is
\beq
c_{\phi+Y}^{SUSY} = 3 + 3Q^2.
\label{cphiY}
\eeq
The criticality condition for the string on \eqref{target} implies  that this central charge should be 
equal to 9. This gives 
\beq
Q=\sqrt{2}\,.
\label{Q}
\eeq

Deformation of the conifold \eqref{deformedconi} translates into adding the Liouville interaction
to the world-sheet sigma model \cite{GivKut}
\beq
\delta L= b\int d^2\theta \, e^{-\frac{\phi + iY}{Q}}\,.
\label{liouville}
\eeq
The conifold singularity at  $b=0$ corresponds to the string coupling constant becoming infinitely large at
$\phi \to -\infty$, see \eqref{strcoupling}. At $b\neq 0$ the Liouville interaction regularize the behavior
of the string coupling preventing the string from propagating to the region of large negative $\phi$.

In fact the $c=1$ non-critical string theory  can also be described in terms of two-dimensional
black hole \cite{Wbh}, which is the ${\rm SL}(2,R)/{\rm U}(1)$ coset WZNW theory 
\cite{MukVafa,GVafa,OoguriVafa95,GivKut} 
at level
\beq
k =\frac{2}{Q^2}\,.
\eeq
This relation implies in the case of the conifold ($Q=\sqrt{2}$)  that 
\beq
k=1,
\label{k=1}
\eeq 
where $k$ is the total level of the Ka\v{c}-Moody algebra in the supersymmetric version (the level
of the bosonic part of the algebra is then $k_b=k+2 =3$). The target 
space of 
this theory has the form of a semi-infinite cigar;  the field $\phi$ associated with the motion along the 
cigar
cannot take large negative values due to semi-infinite geometry. In this description the string
coupling constant at the tip of the cigar is $g_s \sim 1/b$.

\section {Vertex operators and the spectrum}
\label{spectrum}
\setcounter{equation}{0} 

In this section we consider vertex operators for the non-critical string theory on \eqref{target} and 
calculate the string spectrum.

\subsection{Vertex operators}

Vertex operators for  the  string theory on \eqref{target} are constructed in \cite{GivKut}, see also 
\cite{MukVafa,GivKutP}. Primaries of the $c=1$  part for large 
positive $\phi$ (where the target space becomes a cylinder $\mathbb{R}_{\phi}\times S^1$) take the form
\beq
V_{j,m}\approx \exp{\left(\sqrt{2}j\phi + i\sqrt{2}mY\right)}.
\label{vertex}
\eeq
 For the self-dual radius \eqref{Q} (or $k=1$) the parameter $2m$ in Eq. (\ref{vertex}) 
is integer. In the
left-moving sector $2m$ is 
the total momentum plus the winding number along the compact dimension $Y$. For the right-moving sector
we introduce $2\bar{m}$ which is the winding number minus momentum;  then we consider operators with 
$\bar{m}=\pm m$.

The primary operator \eqref{vertex} is related  to the wave 
function over ``extra dimensions'' as follows:
$$V_{j,m} = g_s \Psi_{j,m}(\phi,Y)\,.$$ The string coupling 
\eqref{strcoupling}  depends on $\phi$. Thus, 
\beq
\Psi_{j,m}(\phi,Y) \sim e^{\sqrt{2}(j+\frac{1}{2})\phi + i\sqrt{2}mY}\,.
\eeq
We look for string states with normalizable wave functions over the ``extra dimensions'' which we will 
interpret as hadrons of 4D \ntwo QCD. The 
condition for the string states to have  normalizable wave functions reduces to
 \cite{GivKut}\,
\beq
j\le -\frac12\,.
\label{normalizable}
\eeq
The scaling dimension of the primary operator \eqref{vertex} is 
\beq
\Delta_{j,m} = m^2 - j(j+1) \, .
\label{dimV}
\eeq
Unitarity implies that it should be positive,
\beq
\Delta_{j,m}> 0\,.
\label{Deltapositive}
\eeq

The spectrum of the allowed values of $j$ and $m$ in \eqref{vertex} was determined exactly by using the Kac-Moody algebra
for the coset ${\rm SL}(2,R)/{\rm U}(1)$ in \cite{DixonPeskinLy,Petrop,Hwang,EGPerry,MukVafa}, 
see \cite{EGPerry-rev} for a review. Both discrete and continuous representations were found. Parameters $j$
and $m$ determine the global quadratic Casimir operator and the projection of the spin  on the 3-axis,
\beq
J^2\, |j,m\rangle\, = -j(j+1)\,|j,m\rangle, \qquad J^3\,|j,m\rangle\, =m \,|j,m\rangle
\eeq
where $J^a$  $(a=1,2,3)$ are global ${\rm SL}(2,R)$ currents. We have 

(i) {\em Discrete representations} with
\beq
j=-\frac12, -1, -\frac32,..., \qquad m=\pm\{j, j-1,j-2,...\}.
\label{descrete}
\eeq

(ii) {\em Principal} continuous representations with
\beq
j=-\frac12 +is, \qquad m= {\rm integer} \quad {\rm or} \quad m= \mbox{ half-integer},
\label{principal}
\eeq
where $s$ is a real parameter and

(iii) {\em Exceptional} continuous representations with
\beq
-\frac12 \le j < 0, \qquad m= {\rm integer}.
\label{exceptional}
\eeq

We see that discrete representations include the normalizable states localized near the tip of the cigar,
 while the continuous representations contain
non-normalizable states,\footnote{We will discuss the case $j=-\frac12$
which is on the borderline between normalizable and non-normalizable states in the next subsection.}  see \eqref{normalizable}. This nicely matches our qualitative expectations.

Discrete representations contain
states with negative norm. To exclude the ghost states a restriction for spin $j$ is imposed 
\cite{DixonPeskinLy,Petrop,Hwang,EGPerry,EGPerry-rev} 
\beq
 -\frac{k+2}{2}< j <0\,.
\eeq
Thus, for  our value $k=1$ we are left with only two allowed values of $j$,
\beq
j=-\frac12, \qquad m= \pm\left\{\,\frac12,\, \frac32,...\right\}
\label{j=-1/2}
\eeq
and 
\beq
j=-1, \qquad m= \pm\{\,1, \,2,...\}.
\label{j=-1}
\eeq

Below in this section we will first consider normalizable string states from discrete representations and 
finally, in Subsec. \ref{nonn}, discuss physical interpretation of the continuous representations.

\subsection{Massless baryon}
\label{masslbar}

Our first task now is to rederive the massless baryon $b$ associated with deformations of the conifold complex structure   (see \cite{KSYconifold} and Sec.~\ref{conifold}) within the framework of  the non-critical
Liouville string theory described above.
To this end we consider vertex operators for 4D scalars.
The 4D scalar vertices $V^S$   in the $(-1,-1)$ picture have the
form \cite{GivKut}
\beq
V^S_{j,m}(p_{\mu})= e^{-\varphi}\, e^{ip_{\mu}x^{\mu}}\, V_{j,m}\, ,
\label{tachyon}
\eeq
where $\varphi$ represents bosonized ghosts, and  $p_{\mu}$ is the 4D momentum of the string state.
These states are the lowest components of \ntwo multiplets in four dimensions.
Also, the GSO projection restricts the integer $2m$ for the operator in \eqref{tachyon} to be odd
\cite{KutSeib,GivKut},
\beq
 m=l+\frac12, \qquad |l| = 0,1,2,... 
\label{oddn}
\eeq
The condition for the  state \eqref{tachyon} to be physical is
\beq
\frac{p_{\mu}p^{\mu}}{8\pi T} + \frac{(2l+1)^2}{4} - j(j+1) =\frac12\,,
\label{tachphys}
\eeq
where we used \eqref{dimV} and \eqref{oddn}.
This determines the masses of the 4D scalars, 
\beq
\frac{(M^S)^2_{j,l}}{8\pi T}=-\frac{p_{\mu}p^{\mu}}{8\pi T} = \frac{(2l+1)^2}{4} -\frac12 - j(j+1)\,,
\label{tachyonmass}
\eeq
where the Minkowski 4D metric with the diagonal entries $(-1,1,1,1)$ is used.

Consider the states which are on the borderline between normalizable and non-normalizable, namely, the states with
\beq
j=-\frac12\, .
\label{s=0}
\eeq
For  $l=0$ ($m=\frac12$) and $l=-1$  ($m=-\frac12$) , Eq. \eqref{tachyonmass} gives the lightest states with
\beq
M^S_{j=-\frac12,l=0}= M^S_{j=-\frac12,l=-1}=0.
\eeq
This is our massless baryon $b$ associated with deformations of the complex structure of the conifold (plus
anti-baryon $\tilde{b}$). To confirm
this let us show that it has a logarithmically normalizable wave function in terms of the conifold radial
coordinate $\widetilde{r}$, see \eqref{Sb}.

For $j=-\frac12$  all states \eqref{tachyon} have constant wave functions with respect to the 
Liouville coordinate $\phi$. Thus, the norm of these states is $(\phi_{\rm max}-\phi_{\rm min})$. 
To relate $\phi$ to $\widetilde{r}$
we note that $\phi_{\rm min}$ is determined by the Liouville interaction term \eqref{liouville} which becomes
of the order of unity at this point. This gives
\beq
\phi \sim \log{\widetilde{r}^{\, 2}}\, ,
\label{phitilder}
\eeq
where we used that $\widetilde{r}_{\rm min}^{\,\, 2}=|b|$. In particular, this gives 
$\log{\widetilde{r}^{\,\, 2}_{\rm max}/|b|}$
 for the norm of  the massless state with $l=0$, as expected (see \eqref{Sb}).

\subsection{Massive 4D scalars}
\label{massi}

Now consider the states \eqref{tachyon} with  arbitrary values of $l$ in \eqref{oddn}, still assuming $j=-\frac12$. 
From  \eqref{tachyonmass} we obtain their masses
\beq
(M^S)^2_{j=-\frac12,l} = 8\pi T\,\,l(l+1), \qquad |l|=0, 1,2,...
\label{massivescalars}
\eeq
All these states are logarithmically normalizable with respect to the conifold radial coordinate.

\begin{figure}
\epsfxsize=10cm
\centerline{\epsfbox{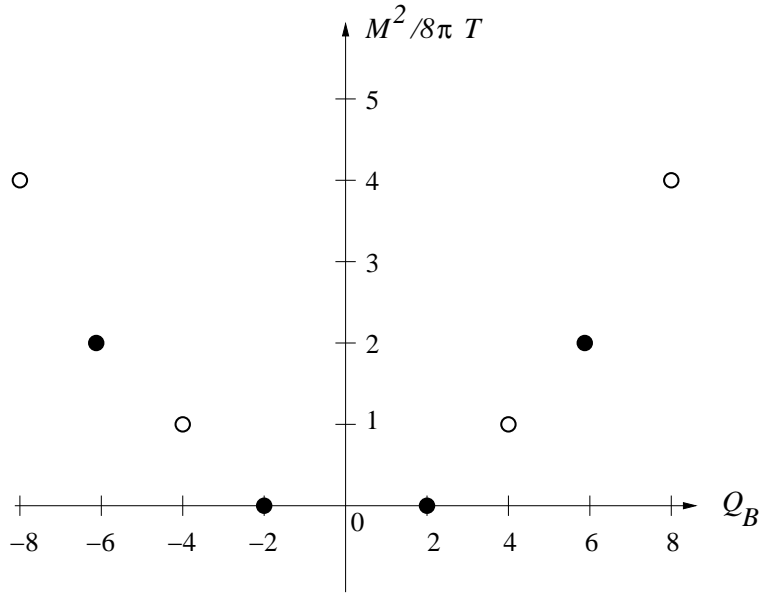}}
\caption{\small  Spectrum of spin-0 and spin-2 states as a function of the baryonic charge. Closed 
and open circles denote  spin-0 and spin-2 states, respectively.
 }
\label{fig_spectrum}
\end{figure}

What are their quantum numbers with respect to the 4D global group \eqref{c+f}? They are all invariant with 
respect to SU(2) factors. To determine their baryon charge note that the U(1)$_B$ transformation of $b$ in 
the Liouville interaction \eqref{liouville} is compensated by a shift of $Y$. Therefore,  $m$ in 
\eqref{vertex}
is proportional to the baryon charge. Normalization is fixed by the massless baryon $b$ which 
has $Q_{B}(b)=2$ at $m=\frac12, \,\, l=0$. This implies 
\beq
 Q_B(V_{j,m}) = 4m.
\label{n-baryon}
\eeq
We see that the momentum $m$ in the compact $Y$ direction is in fact the baryon charge of a string state.
In particular, 4D scalar states \eqref{massivescalars} are all  baryons for positive $l$ and
anti-baryons for negative $l$  with $$Q_B= 4l+2\,.$$ The masses of 4D scalars as a function of the 
baryonic charge are shown in Fig.~\ref{fig_spectrum}.

To conclude this subsection let us note that the second allowed value of $j$, $j=-1$ in \eqref{j=-1},
is excluded by the GSO projection which selects only half-integer values of $m$ for states \eqref{tachyon}, see \eqref{oddn}.

Note also that the 4D scalar states found above are the lowest components of \ntwo multiplets. Other 
components can be restored by virtue of 4D \ntwo supersymmetry.

 \subsection{Spin-2 states}
 \label{s2s}

At the next level we consider  4D spin-2 states coming from space-time ``gravitons.'' 
The corresponding vertex operators are given by
\beq
V^G_{j,m}(p_{\mu})= \xi_{\mu\nu}\psi^{\mu}_L\psi^{\nu}_R\,e^{-\varphi}\, e^{ip_{\mu}x^{\mu}}\, V_{j,m}\, ,
\label{graviton}
\eeq
where $\psi^{\mu}_{L,R}$ are the world-sheet superpartners to 4D coordinates $x^{\mu}$; moreover, 
$\xi_{\mu\nu}$ is the polarization tensor.

The condition for these  states  to be physical takes the form
\beq
\frac{p_{\mu}p^{\mu}}{8\pi T} + m^2 - j(j+1) = 0\,.
\label{gravphys}
\eeq

The GSO projection selects now $2m$ to be even, $m=l$, $\, |l|=0, 1,2,...$ \cite{GivKut},
thus we are left with only one allowed value of $j$, $j=-1$ in \eqref{j=-1}. Moreover, the value
$m=l=0$ is excluded.
This leads to the following expression  for the masses of spin-2 states:
\beq
(M^G)^2_{j,l} = 8\pi T\,l^2, \qquad |l|=1,2,... .
\label{gravitonmass}
\eeq
We see that all spin-2 states are massive. This confirms the result in \cite{KSYconifold} that 
no massless 4D graviton appears in our theory. It also matches the fact that our ``boundary'' theory, 4D 
\ntwo QCD, is defined in flat space without gravity.

All states with masses \eqref{gravitonmass}  are baryons for  $l>0$ and
anti-baryons for  $l<0$, with the baryon charge $Q_B=4m=4l$. The masses of 4D spin-2 states as a 
function of the baryonic charge are shown in Fig.~\ref{fig_spectrum}.

We expect that all 4D states with a given 
baryon charge considered in this and previous subsections 
are the lowest states of the Regge trajectories linear in 4D spin.

\subsection{Non-normalizable states}
\label{nonn}

In \cite{KSYconifold} the continuum spectrum of non-normalizable  states was
 interpreted as unstable string states. Since the
Liouville coordinate $\phi$ is related to the radial coordinate $\widetilde{r}$ on the conifold (see
\eqref{phitilder}) the modes with $j> -\frac12$ are power non-normalizable on the conifold.
The conifold radial coordinate $\widetilde{r}$ has the physical interpretation of a distance from
the string axis in 4D space, see \eqref{tilder} and \eqref{w}. Therefore, the wave functions of 
the non-normalizable states are saturated at large distances from the 
vortex-string axis  in 4D. 

These states are {\em not} localized on the string. The infinite norm of these 
states should be interpreted as an instability. Namely, these states decay into 
massless bifundamental quarks inherent to the Higgs branch of 
our four-dimensional \ntwo QCD. This instability is present already at the perturbative level, see Sec.~\ref{worldsheet} and 
\cite{KSYconifold}. 

Our vortex string has a conceptual difference  compared to
fundamental string.  In compactifications  of fundamental string  
{\em all} states present in four dimensions are string states.
The string theory for the vortex strings of \cite{SYcstring,KSYconifold} is different. The string 
states describe 
only non-perturbative physics at strong coupling, such as mesons and baryons. The perturbative 
massless moduli 
states seen at week coupling are not described by this theory.
 In particular, the Higgs branch (and associated massless bifundamental quarks)
 found at weak coupling can be continued to the strong coupling; they persist there. It can
intersect other branches, but cannot disappear (for quarks with the vanishing mass terms)
\cite{APS}.

One class of non-normalizable unstable modes is given by the exceptional continuous representation
\eqref{exceptional}. For $|m|=1,2,...$  the continuous spectra parametrized by $j$ in \eqref{exceptional} 
start from the thresholds given by 
masses \eqref{gravitonmass}.

Another class of  unstable string states corresponds to the principal representation  given by 
complex values of $j$
\beq
j=-\frac12 +is,
\eeq
see \eqref{principal}. These states have  continuous spectra parametri\-zed by $s$.
The parameter $s$ has a clear-cut interpretation of a momentum along the Liouville direction.
Therefore we interpret these states as decaying modes of the string states interacting with the perturbative
bifundamental quarks rather than the hadronic states of \ntwo QCD.

Much in the same way as for the exceptional representation the principal contunuous spectra for 
half-integer $m$  start from thresholds given by masses \eqref{massivescalars}.
Using this picture we are led to conclude that the continuous spectra contain  multi-particle states
formed by a given baryon and a number of emitted bifundamental quarks with zero total baryonic charge.
This issue needs a future clarification.

\section {Physical interpretation of string states}
\label{necklace}
\setcounter{equation}{0} 

In this section we reveal a physical interpretation of all baryonic  states found in the previous section
 as  monopole ``necklaces.''

\begin{figure}
\epsfxsize=10cm
\centerline{\epsfbox{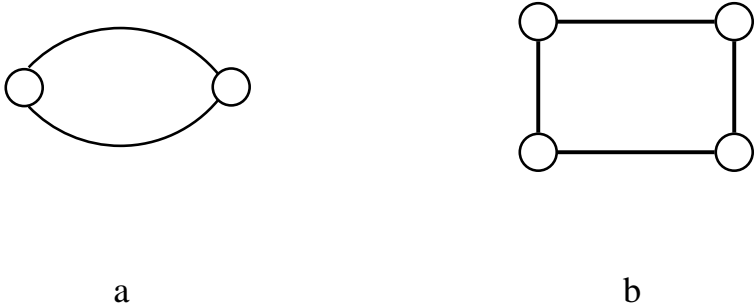}}
\caption{\small  Examples of the monopole ``necklace'' baryons: a) Massless $b$-baryon with $Q_B=2$; 
b) Spin-2 baryon with $Q_B=4$.
Open circles denote  monopoles. }
\label{baryons}
\end{figure}

Consider first the weak coupling domain $g^2\ll 1$ in  four-dimensional \ntwo QCD. It is in the Higgs phase:  $N$ squarks  condense. Therefore,   non-Abelian 
vortex strings confine monopoles. However, 
the monopoles cannot be attached to the string endpoints. In fact, in the U$(N)$ theories confined  
 monopoles 
are  junctions of two distinct elementary non-Abelian strings \cite{T,SYmon,HT2} (see \cite{SYrev} 
for a review). As a result,
in  four-dimensional \ntwo QCD we have 
monopole-antimonopole mesons in which the monopole and antimonopole are connected by two confining strings.
 In addition, in the U$(N)$  gauge theory we can have baryons  appearing as  a closed 
``necklace'' configurations of $N\times$(integer) monopoles \cite{SYrev}. For the U(2) gauge group the 
lightest baryon presented by such a ``necklace'' configuration  
consists of two monopoles, see Fig.~\ref{baryons}.

Moreover, the monopoles acquire quantum numbers with respect to the global symmetry group (\ref{c+f}). To see that this is the case note that in the world-sheet
theory on the vortex string the confined monopole is seen as a kink interpolating between two distinct vacua
(i.e.  distinct elementary non-Abelian strings) in the corresponding 2D sigma model \cite{T,SYmon,HT2}. 
At the same time, we know that the sigma model kinks at strong coupling are described by the $n^P$ and $\rho^K$  fields \cite{W79,HoVa}
(for the sigma model described by \eqref{wcp} it was shown in \cite{SYtorkink}) and therefore transform in 
the fundamental representations\,\footnote{Strictly speaking, 
to make  both bulk monopoles and world-sheet kinks well defined as localized objects 
one should introduce an infrared regularization,
say, a small quark mass term. When we take the limit of the zero quark masses, the kinks become massless
and smeared all over the closed string. However their global quantum numbers stay intact.} of 
two non-Abelian factors in (\ref{c+f}).

As a result, the monopole  baryons in our case can be singlets, triplets, bifundamentals or form higher
representations 
of both SU(2) global groups in (\ref{c+f}). With respect
to the baryonic U(1)$_B$ symmetry in (\ref{c+f}) the 
monopole baryons can have charges
\beq
|Q_{B}({\rm baryon})|=0,\,1,\,2\,...,
\label{Bbaryons}
\eeq
 see (\ref{repsnrho}). In particular, non-zero baryonic charge is associated with the $\rho$ kinks.
In the U(2) gauge theory the monopole ``necklace'' can be formed by even number of monopoles.

All these non-perturbative stringy states are heavy at weak coupling, with mass of the
order of $\sqrt{\xi}$, and therefore can decay into screened quarks which are lighter and, eventually, into
massless bifundamental screened quarks.

Now, we pass to the self-dual point $\beta=0$ in the strong coupling region. As was already discussed 
all string states found in Sec.~\ref{spectrum}  have non-zero
baryon charge $Q_B=4m$, see Eq. \eqref{n-baryon}. 
The lightest state (the massless $b$ state) has $Q_B=2$. It can be formed 
by minimum two monopoles, see Fig.~\ref{baryons}. The spin-2 massive state \eqref{graviton} 
with $m=1$ can be formed by the  monopole ``necklace'' with minimum four monopoles. 

All stringy monopole ``necklace'' baryons found in 
Sec.~\ref{spectrum} are singlets with respect to two SU(2) factors in \eqref{c+f}. They are metastable and
can decay into pairs of 
massless bifundamental
quarks in the singlet channel with the same baryon charge.
The metastability of stringy baryons  on the string side is reflected  in the logarithmic divergence of 
their norm and the presence of continuous spectra.
Detailed studies of the non-perturbative Higgs branch formed by VEV of massless $b$ and   
interactions of stringy baryons with massless 
bifundamental quarks are left for future work.

\section {Conclusions}
\label{conclusions}
\setcounter{equation}{0} 

Previously  we observed that non-Abelian vortex strings supported in four-dimen\-sional 
\ntwo supersymmetric QCD with the U(2) gauge group
and $N_f=4$ flavors of quark hypermultiplets  can represent critical superstrings in ten-dimensional
target space $\mathbb{R}^4\times Y_6$ where $Y_6$ is a noncompact Calabi-Yau manifold.
This can be called ``reverse holography."
Indeed, we start from a well-defined four-dimensional Yang-Mills theory and, analyzing the vortex strings 
it supports, add six extra dimensions -- the moduli of the string world-sheet theory -- which relates our 
construction (at the critical value of the coupling constant $g^2=4\pi$ corresponding to $\beta=0$)
to a critical string theory on a six-dimensional conifold.

In this paper the mass spectrum of the string states is determined using
equivalent formulation in terms of $c=1$ non-critical string theory with the Liouville field. The
 string  states {\em per se}
are identified with the hadronic states in the four-dimensional theory. Since the extra-dimensional space 
is not compact the above identification becomes possible because our string theory 
is holographic: non-trivial physics is 
``projected'' to 4D ``boundary''. This behavior is typical for LSTs. The reason for holography is 
that the string coupling constant 
exponentially falls-off at large values of the non-compact Liouville coordinate (see \eqref{strcoupling}),
and the bulk physics becomes trivial and decouples. 

Of course, the holography of our string theory is expected since our starting point was 4D \ntwo QCD.
Holography ensures the presence of normalizable string states 
localized in 4D space which we identified as hadrons of \ntwo QCD. Also, we qualitatively interpret 
non-normalizable states as decay modes of the  hadronic states.

The Higgs branch existing in our four-dimen\-sional 
\ntwo  QCD leads to  strictly massless quark hypermultiplets, which are evident at weak coupling and survive the transition to strong coupling. This implies a peculiar behavior in the infrared. In particular,
the masses of the string states derived in Sec. \ref{massi} and Sec. \ref{s2s} are in fact the endpoints of the branch cuts. Note, however, that exactly the same situation would take place in QCD
with massless quarks giving rise to massless pions. Say, every lowest-lying state with the given baryon number would represent the beginning of a branch cut. The pion mass could be lifted by an arbitrarily small perturbation. Disentangling infrared effects of the Higgs branch from physics of the critical string under consideration will be the subject of a subsequent work. 

\section*{Acknowledgments}

The authors are grateful to Peter Koroteev and Xi Yin  for very useful and 
stimulating discussions.
This work  is supported in part by DOE grant DE-SC0011842. 
The work of A.Y. was  supported by William I. Fine Theoretical Physics Institute  at the  University 
of Minnesota, and by Russian State Grant for
Scientific Schools RSGSS-657512010.2. The work of A.Y. was supported by Russian Scientific Foundation 
under Grant No. 14-22-00281.

\end{document}